\documentstyle[aps,epsf]{revtex}
\tighten
\begin{document}

\title{The bound $\mu^+\mu^-$ system}

\author{U. D. Jentschura
\footnote{Electronic address: ulj@nist.gov}, G. Soff}
\address{Institut f\"ur Theoretische Physik, TU Dresden,
01062 Dresden, Germany }

\author{V. G. Ivanov}
\address{Pulkovo Observatory, 196140, St. Petersburg, Russia}

\author{S. G. Karshenboim\footnote{Electronic address:
sgk@onti.vniim.spb.su}}
\address{D. I.  Mendeleev Institute for Metrology, 
198005, St.~Petersburg, Russia,\\
Max--Planck--Institut f\"{u}r Physik komplexer
Systeme, Bayreuther Stra\ss e 40, 01069 Dresden, Germany}

\maketitle

\begin{abstract}
We consider the hyperfine structure, the atomic spectrum
and the decay channels of the bound $\mu^+\mu^-$ system 
(dimuonium).
The annihilation
lifetimes of low-lying atomic states 
of the system lie in the $10^{-12}\,{\rm s}$
range. The decay rates could be measured 
by detection of the decay 
products (high energy photons or electron-positron pairs).
The hyperfine structure splitting
of the dimuonic system and its decay rate are influenced by  
electronic vacuum 
polarization effects in the far time-like asymptotic region.
This constitutes a previously unexplored kinematic regime.
We evaluate next--to--leading order
radiative corrections to the decay rate of low-lying atomic states.
We also obtain order $\alpha^5\,m_{\mu}$ corrections to the 
hyperfine splitting of the $1S$ and $2S$ levels. 
\end{abstract}

\pacs{ PACS numbers 36.10.Dr, 12.20.Ds, 31.30.Jv}

\section{INTRODUCTION}
\label{intro}

The bound system consisting of two muons (dimuonium)
can be produced in 
heavy nuclei inelastic scattering at high energies
and in particle decays. 
The decay of the neutral $\eta^0$ meson into dimuonium
has been investigated theoretically 
by L. Nemenov \cite{nemenov}
($\eta^0 \to {\sl dimuonium} + \gamma$).
The formation of dimuonium in pion--proton collisions
($\pi^{-} + p \to {\sl dimuonium} + n$) and 
by photons on nuclei ($\gamma + Z \to {\sl dimuonium} +
Z$) has been discussed in \cite{bilenkii}.
For the direct production of dimuonium in muon--antimuon
collisions, considerable experimental difficulties
associated with slow muon beams would have to be overcome.
Another possible pathway for the production of the system,
which we do not discuss in any further detail here, is the 
$e^+e^-$-annihilation 
(near or above the $\mu$ threshold). 
Dimuonium, once produced, undergoes
atomic decay (into energetically lower atomic states) and 
annihilation decay (into electrons and photons).
Because the annihilation products are
hard photons and relativistic
electron--positron pairs, the decays could be 
investigated experimentally by established methods 
of particle physics. 

In this work, we devote special attention to the
annihilation decay rates of low-lying atomic states.
The decay rate of the $S=1$ ortho states of dimuonium
has been evaluated by J. Malenfant in pioneering investigations
\cite{malenfant1,malenfant2}.
The name ``dimuonium'' has been proposed in \cite{malenfant2}.
This work can therefore be regarded as a continuation
of the earlier investigations on the system.
We try to refine the analysis 
of the ortho--states, and we investigate in addition 
the $S=0$ para states. We obtain
corrections to the decay rate in next--to--leading order.

In this work, we also investigate the hyperfine structure
in next--to--leading order. 
We further discuss briefly the atomic spectrum
of dimuonium.
The contribution of electronic vacuum
polarization to the hyperfine structure
and to the decay rate of ortho states
lies in the far time-like asymptotic 
region. Its
observation would constitute a test of 
QED in this previously unexplored 
kinematic regime.
Due to the small length scale
of the system, the hyperfine structure
of the dimuonic atom is influenced by the effect of hadronic
vacuum polarization (at the level of one part in $10^3$). 

Among the exotic atomic systems, some attention has 
recently been devoted to 
pionium ($\pi^+\pi^-$-system). The spectrum and
decay channels of pionium 
have been discussed extensively (see e. g. \cite{pion1,pion2,pion3}).
The production of pionium has been observed recently 
\cite{pipi2}. 
The formation of bound atomic systems in particle decays 
has been observed earlier. Positronium has been detected
in the decay $\pi^0\to {\rm Ps}+\gamma$ \cite{posgamdecay}. 
The bound system of a pion and a muon ($\pi\mu$) has been 
observed in the 
decay $K^0_L\to (\pi\mu\,{\rm atom})+\nu$ \cite{pimudecay}. 

In contrast to 
the short--lived $\pi^{+} \pi^{-}$ system, 
where $c \cdot \tau$ is in the range of $1\,\mu{\rm m}$,
$c \cdot \tau$ is in the range of
$1\,{\rm mm}$ for dimuonium, so that the atom could leave the 
beam target after production. In any of the cases discussed
the production rate is expected to be proportional
to the atomic wave function at the origin 
$|\psi_{nl}(0)|^2=\alpha^3\,m_\mu^3\,\delta_{l0}/[8\,\pi\, n^3]$
(we use relativistic units 
in which $ \hbar  = c = 1$ and $\alpha=e^2$). 
Hence only low-lying 
states with zero orbital moment 
($S$ states) are expected 
to be produced. 

This paper is organized as follows.
We first study the spectrum of the system including radiative
corrections (Lamb shift, Section \ref{spectrum}). 
We then proceed to an investigation of
the hyperfine structure in Section \ref{hfs}. 
As the highest
accuracy can presumably be obtained 
in decay rate measurements, 
we devote special attention to the decay
channels which are investigated in detail
in Sections \ref{secOM} and \ref{secPM}.

\section{THE ATOMIC SPECTRUM OF DIMUONIUM}
\label{spectrum}

\begin{figure}[htb]
\centerline{\mbox{\epsfysize=10.5cm\epsffile{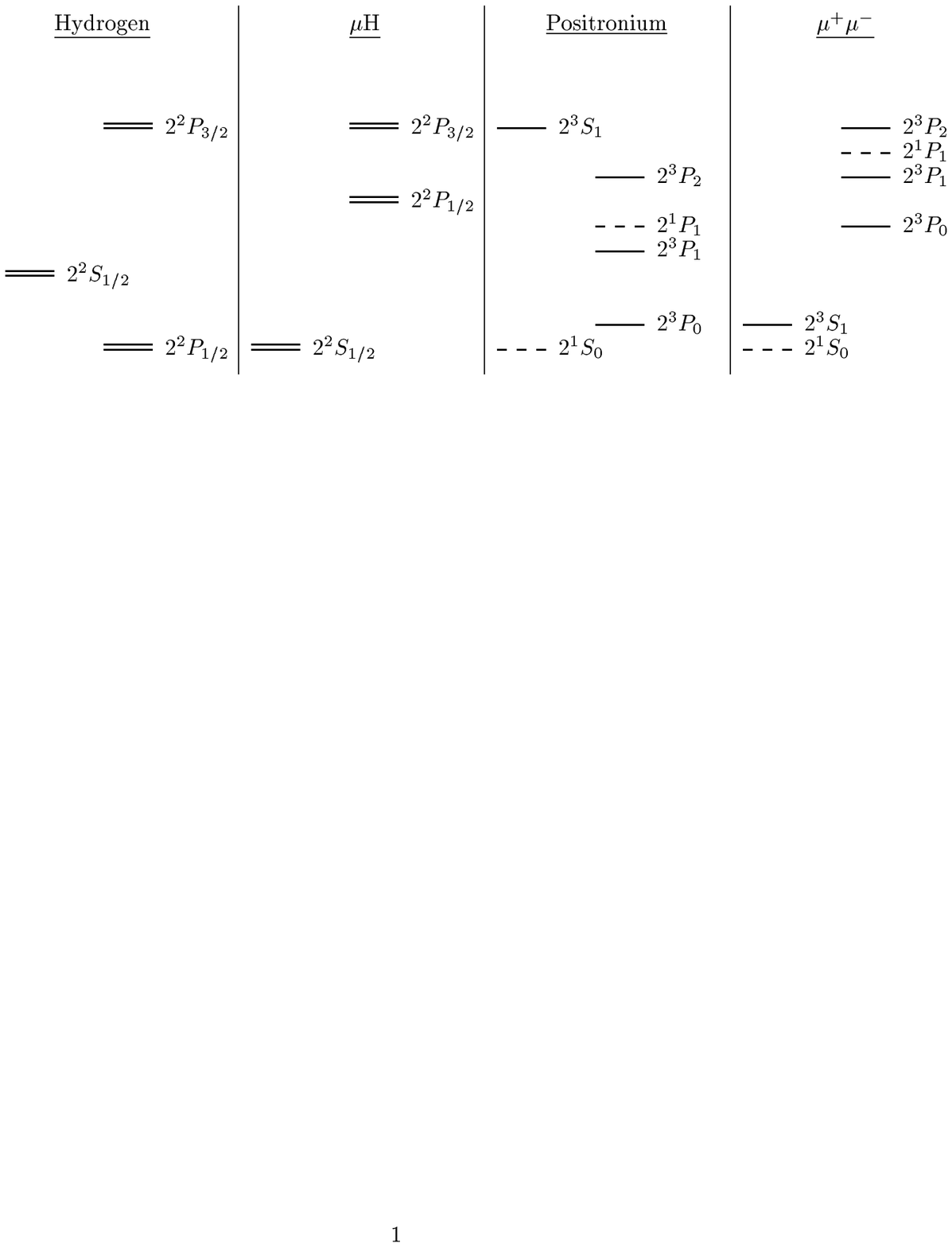}}}
\caption{\label{oviewspec} Overview over the spectrum of
$n=2$ levels in hydrogen, muonic hydrogen ($\mu$H), positronium and 
dimuonium ($\mu^{+}\mu^{-}$). 
The double lines denote the hyperfine structure
splitting of the levels in hydrogen and muonic hydrogen. Dashed lines
denote $S=0$-states, full lines denote $S=1$-states.}
\end{figure}

The dimuonic atom is an analogue of the hydrogen, muonic hydrogen and 
positronium atoms. The main feature 
of muonic hydrogen ($\mu$H) which differs from properties of hydrogen 
is the order of magnitude of the Lamb shift relative to
the fine structure splitting (see e. g. \cite{IV}). 
The fs splitting in $\mu$H and $H$ (also in positronium and
dimuonium) is of 
the order of $\alpha^4\,m_r$,
where $m_r$ denotes the reduced mass of
the system, which is roughly equal to the mass 
of the lighter particle
in a system with a heavy nucleus. 
In hydrogen the Lamb shift of the order of $\alpha^5\,m_r$, 
whereas in muonic systems,
the Lamb shift is of the order of $\alpha^3\,m_r$,
because of the large effect of vacuum polarization.
The above considerations are
parametrically true for all states, but numerical coefficients
turn out to be largest for $S$ states. Because the coefficients 
to the Lamb shift are negative (attractive Uehling potential),
$S$ states are significantly lowered in energy in muonic systems.
In Fig. \ref{oviewspec} we present an overview over the spectrum
of hydrogen, $\mu$H, positronium and $\mu^{+}\mu^{-}$. 

In both hydrogen and $\mu$H, the hyperfine 
structure is a smaller effect of order $\alpha^4\,(m_r^2/M)$
($M$ denotes the mass of the nucleus). 
This is different in positronium.
Positronium has a more complicated spectrum than the hydrogen 
atom. Due to the larger magnetic moment of the ``central'' 
particle (and its smaller mass), and due to virtual pair annihilation
processes, the effects which are responsible for the fine and hyperfine 
structure and for the Lamb shift have the same order of magnitude: 
$\alpha^4\,m_r$. For positronium as well as for hydrogen, 
the fine structure is defined as the separation
of levels with the same principal and angular orbital 
momentum quantum numbers. 

In positronium, 
different quantum numbers have to be used for the 
classification of levels than in 
an atom with a heavy nucleus. For a system with a heavy
nucleus, the total electron moment ${\bbox j}={\bbox l}+{\bbox s}$ 
is conserved within the central field approximation. 
States are classified as 
$nljF$. In positronium the total spin 
${\bbox S}={\bbox s}_1+{\bbox s}_2$ 
is an exactly conserved quantity. States are classifyed as $nlSF$
(denoted as $n^{2\,S + 1}l_F$), where
$S=1$ for an ortho and $S=0$ for para states. 
In all systems considered we denote the total angular atomic moment by $F$.
Although the atomic spectrum of dimuonium
is determined by a large Lamb shift of order 
$\alpha^3\,m_r$ (as is the case for $\mu$H), the 
classification of levels is like in positronium (quantum numbers $nlSF$). 

There exists no close analogy of dimuonium with pionium, because
pionium containes only spinless particles, 
and the spectrum includes neither a fine nor a hyperfine structure.
Due to the strong interaction, the hadronic pionium
system lives only a very short time,
and the atomic structure cannot be investigated (because 
the decay rate exceeds the rate of atomic transitions). By contrast, 
for $P$-states in dimuonium,
the atomic transition $nP\to n'S$ ($n' < n$) dominates
over the annihilation decay rate.
The lifetime of the atomic 
levels is considered in section \ref{hfs}. 

The main contribution to the Lamb shift in dimuonium is caused by
the one--loop electronic vacuum polarization. Due to the small spatial
separation of the two muons in the $\mu^{+}\mu^{-}$ system, the
charge of the two particles, which is larger than the
effective charge to be observed at larger distance, is less screened
and causes a deviation
of the energy levels of the order $\alpha^3\,m_r$. For hydrogen, the vacuum
polarization enters at the order of $\alpha^5\,m_r$, 
but for dimuonium, two
powers of $\alpha$ are compensated for by the muon-electron mass
ratio (or, by employing a coordinate space picture, by the smaller
separation of the particles). 

\begin{figure}[htb]
\centerline{\mbox{\epsfysize=4.5cm\epsffile{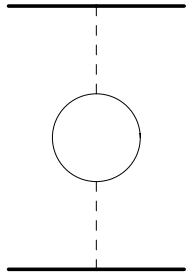}}}
\caption{\label{vacpol} (Electronic) vacuum polarization insertion
in the Coulomb photon (main contribution to the 
Lamb shift in muonic
systems). The dashed line denotes a Coulomb photon. 
Bold fermionic lines denote 
muons, thin lines denote electrons and positrons.} 
\end{figure}

We now turn to a brief discussion of the atomic levels of dimuonium.
The effect of electronic vacuum polarization on the
spectrum is investigated. Any vacuum polarization insertion in the photon line
can be represented as a substitution of the form
\begin{equation}
\frac{1}{q^2+i\,\epsilon}\to \frac{\alpha}{\pi}\int_{s_0}^\infty
{ds\,\,\rho(s)\,\frac{1}{q^2-s+i\,\epsilon}}
\end{equation}
in the photon propagator. The 
integration has to be performed from the 
threshold $s_0$ of pair production of the
loop particle. For electronic vacuum polarization,
the spectral function is given by (see for instance
\cite{IZ}, p. 323)
\begin{equation}
\label{defrho}
\rho(s) = \frac{1}{3\,s}\,\sqrt{1-\frac{4\,m_e^2}{s}}\,
  \left(1 + \frac{2\,m_e^2}{s}\right)\,.
\end{equation}
The substitution 
\begin{equation}
v^2 = 1 - \frac{4\,m_e^2}{s}
\end{equation}
brings the one-loop vacuum polarization integral into the 
form \cite{Schwinger}
\begin{equation} 
\label{vpe}
\frac{1}{q^2+i\,\epsilon}\to \frac{\alpha}{\pi}\int_0^1
dv\,\frac{v^2(1-v^2/3)}{1-v^2}\,\frac{1}{q^2-\lambda^2+i\,\epsilon}\,.
\end{equation}
For space-like momentum transfer,
this can be Fourier-transformed into coordinate space 
and yields the Uehling potential,
\begin{equation}
\label{uehl}
V_U({\bbox r}) = \frac{\alpha}{\pi}\int_0^1
dv\,\frac{v^2(1-v^2/3)}{1-v^2}
\,\exp\left(-\lambda\,r\right)\,
\left[\frac{-\alpha}{r}\right]\,.
\end{equation}
We have introduced the notation,
\begin{equation}
\label{deflambda}
\lambda = \frac{2\,m_e}{\sqrt{1-v^2}}\,.
\end{equation}
The correction to the energy due to the diagram in Fig. \ref{vacpol}
is the main contribution to the Lamb shift. The evaluation of the matrix
element of the Uehling potential on the non-relativistic wave functions
leads to the results listed in Table \ref{alpha3}. The Rydberg constant
for the dimuonic atom is given by
\begin{equation}
E_0 = \frac{\alpha^2 \, m_r^2}{2} = \frac{\alpha^2 \, m_\mu}{4} = 
1406.6133(5) \, {\rm eV},
\end{equation}
using the recommended values for $\alpha^{-1} = 137.0359895(61)$ and
$m_\mu = 105.658389(91) \, {\rm MeV}$ \cite{cohentaylor1,pdg96}. 
The approximate
Lamb shift values are (neglecting higher order corrections corrections
which are estimated to be suppressed by 
an additional factor of $\alpha/\pi$):
\begin{equation}
{\cal L}(1S) = -0.49 \, {\rm eV},\quad
{\cal L}(2S) = -0.058 \, {\rm eV},\quad
{\cal L}(2P) = -0.0014 \, {\rm eV}.
\end{equation}
In contrast to muonic hydrogen ($\mu$H), 
dimuonium is a purely leptonic
system. Therefore its spectrum is not influenced 
by nuclear structure effects.
 
\section{HYPERFINE STRUCTURE OF DIMUONIUM}
\label{hfs}

\begin{figure}[htb]
\centerline{\mbox{\epsfxsize=10.0cm\epsffile{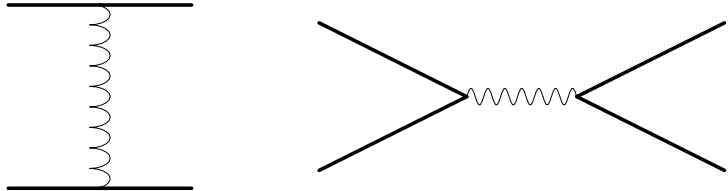}}}
\caption{\label{lowhfs} Transverse photon exchange diagram and
(time-like) photon annihilation diagram. Both of the diagrams
contribute to the hyperfine structure of $S$ states in leading
order. The zig-zag line denotes a transverse photon, the wavy 
line denotes the full photon propagator.}
\end{figure}

The fine and hyperfine structure in dimuonium and 
positronium in lowest order $\alpha^4\,m$ are given 
by the following formula (see for details 
textbooks \cite{IV,IZ,BS} and articles \cite{PosFS3,PosFS1,PosFS2}),  
\begin{equation}  \label{pos4}
E_{\rm Ps}(nlSF) = -\frac{\alpha^2}{4\,n^2}\, m + \alpha^4\, m \,
\bigg[\frac{11}{64\,n^4}+
\frac{\delta_{S,1}}{n^3} \, \left(\frac{7}{6} \, \delta_{l,0} 
+ \frac{1 - \delta_{l,0}}{4(2 \, l + 1)} B_{j,l} \right) 
- \frac{1}{2\,n^3\,(2l+1)} \bigg],
\end{equation}
where
\begin{equation}
B_{j,l} = \left\{
\begin{array}{rl}
\frac{3\,l+4}{(l+1)\,(2\,l+3)} & 
\mbox{for} \quad F=l+1, \\ 
& \\ 
-\frac{1}{l(l+1)} & 
\mbox{for} \quad F=l,\\ 
& \\ 
-\frac{3l-1}{l\,(2\,l-1)} & 
\mbox{for} \quad F=l-1\,.
\end{array} \right. 
\end{equation}
Here $m$ is the mass of the
constituent particle, i. e. $m = m_e$ for positronium, 
$m = m_\mu$ for dimuonium.
Note that the above formula determines the position of energy levels in
positronium up to the order of $\alpha^4$. Radiative corrections in
Ps enter at the order of $\alpha^5$,
at the order of $\alpha^3$ in dimuonium. 
However, for the fine and hyperfine structure of
dimuonium in lowest order, the above formula remains valid. 
The results for the hyperfine structure order in dimuonium 
in leading order are
\begin{eqnarray}
E_{\rm hfs}^{(0)}(nS) = E_F/n^3, \quad \mbox{where} \quad 
E_F = \frac{7}{12}\, \alpha^4 \, m_{\mu} = 0.175 \,{\rm eV} = 
4.23 \cdot 10^8\,{\rm MHz}\,.
\end{eqnarray}
The hyperfine structure interval 
$E_{\rm hfs}^{(0)}(nS)$ arises from the 
exchange and annihilation diagrams
(see Fig. \ref{lowhfs}), which 
contribute $4/7\,E_F$ and $3/7\,E_F$ to
the Fermi energy $E_F$, respectively. 
The triplet $n {^3}S_1$ states 
are energetically higher than the
singlet $n {^1}S_0$ states. For $P$ states, 
the $\alpha^4$ corrections imply
the following order of the spectrum,
in decreasing energy:
$n {^3}P_2$, $n {^1}P_1$, $n{^3}P_1$, $n {^3}P_0$
(we follow here the usual spectroscopic 
notation $n^{2S+1}l_F$). 

\begin{figure}[htb]
\centerline{\mbox{\epsfysize=18.0cm\epsffile{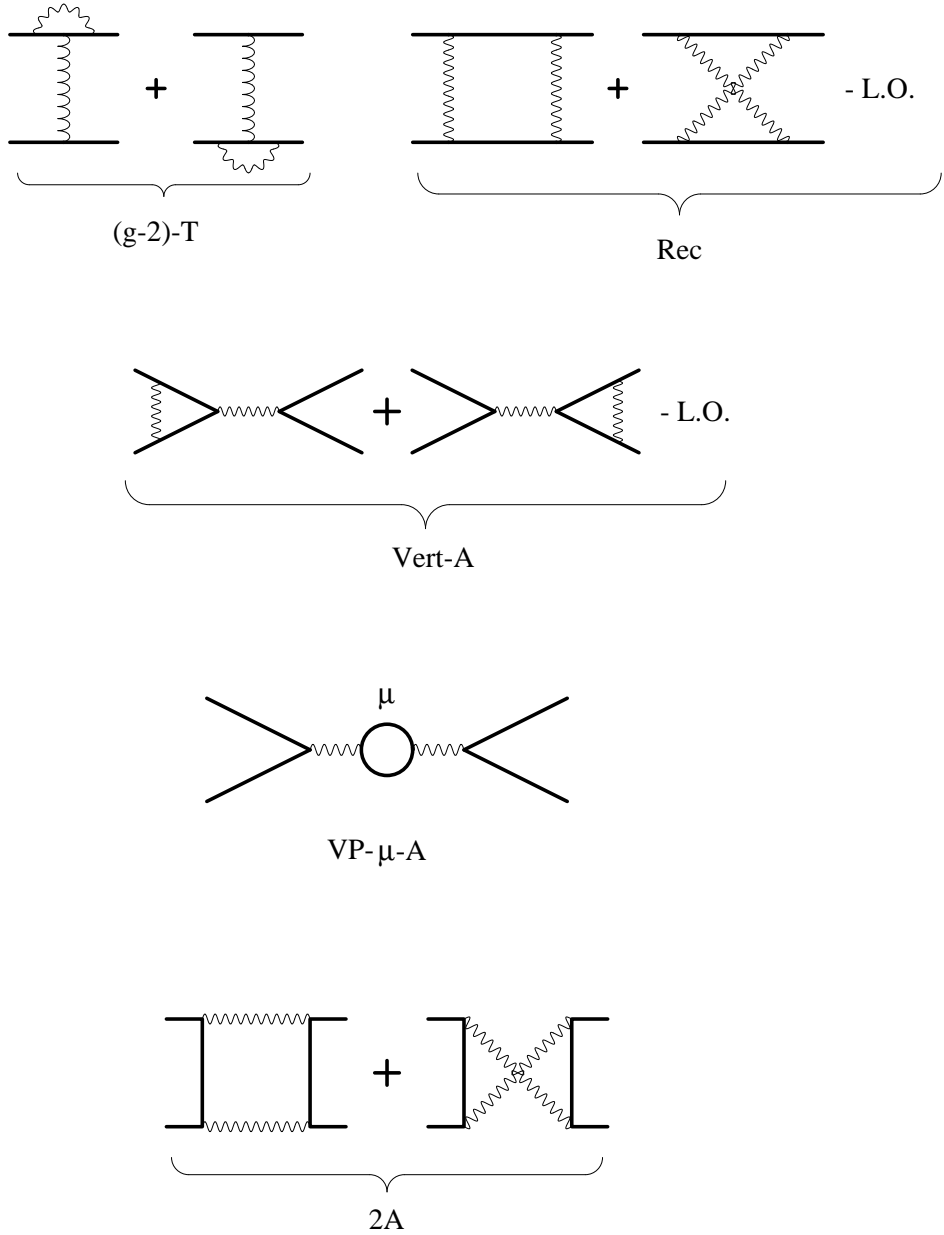}}}
\caption{\label{known} Known corrections up to the 
order $\alpha/\pi\,E_F$ to the hyperfine splitting
of dileptonic systems (positronium, dimuonium). 
Diagrams are explained in the
text. For the dimuonic atom, the corrections depicted here
contribute to hfs, but there are additional terms
specific to the dimuonic system which need
to be taken
into account. The subtraction of lower order (L.O.) contributions is
necessary for some of the diagrams in order to prevent double
counting. The bold fermionic lines denote muons. The direction of time
in all diagrams is from left to right.}
\end{figure}

We now turn to the evaluation of radiative corrections to the
hyperfine structure splitting for $S$ states of order
$\alpha/\pi\,E_F$ 
(corresponding to $\alpha/\pi\,E^{(0)}_{\rm hfs}$). 
Some contributions are the same for positronium and
dimuonium, some are specific to the dimuonic system. 
The contributions which are alike for both systems
are depicted in Fig. \ref{known}.
These contributions originate
from the anomalous magnetic moment of the electron in the
transverse photon diagram (denoted as ``(g-2)-T''), from recoil corrections
(denoted by ``Rec''), from the vertex correction to the annihilation
diagram (denoted by ``Vert-A''), from the muonic vacuum polarization
(``VP-$\mu$-A''), 
and from box-diagrams corresponding to two-photon annihilation
(denoted by ``2A''). 
In this paper, due to the multitude of corrections considered,
we prefer to denote the contributions by 
short abbreviations, which we hope are self--explanatory,
rather then single letters, in order to enhance the
clarity of presentation. The 
results for the above corrections are listed in Table \ref{alpha5}
(first 5 contributions, above the separating line).
The sum of corrections to the Fermi energy of order
$\alpha/\pi\,E_F$ amounts to \cite{IZ,KK}
\begin{equation}  \label{pos5}
\Delta E_{\rm Ps}(nS) = \frac{\alpha}{\pi} \left[ -\frac{32}{21} -
\frac{6}{7} \ln 2 + \frac{3}{7}\,\pi\,i \right] \, \frac{E_F}{n^3} \,,
\end{equation}
where the imaginary part is entirely due to 
paradimuonium. It corresponds to the two-photon decay of the para
system \cite{PosFS1,2phot}
\begin{equation}  
\label{Tpara}
\frac{1}{\Gamma^{(0)}(n {^1}S_0)} =
\tau^{(0)}(n^1S_0) =  \frac{2\,n^3}{\alpha^5 \, m_\mu} =
n^3 \cdot 0.6021 \cdot 10^{-12}\,{\rm s}\,. 
\end{equation}

\begin{figure}[htb]
\centerline{\mbox{\epsfysize=18.0cm\epsffile{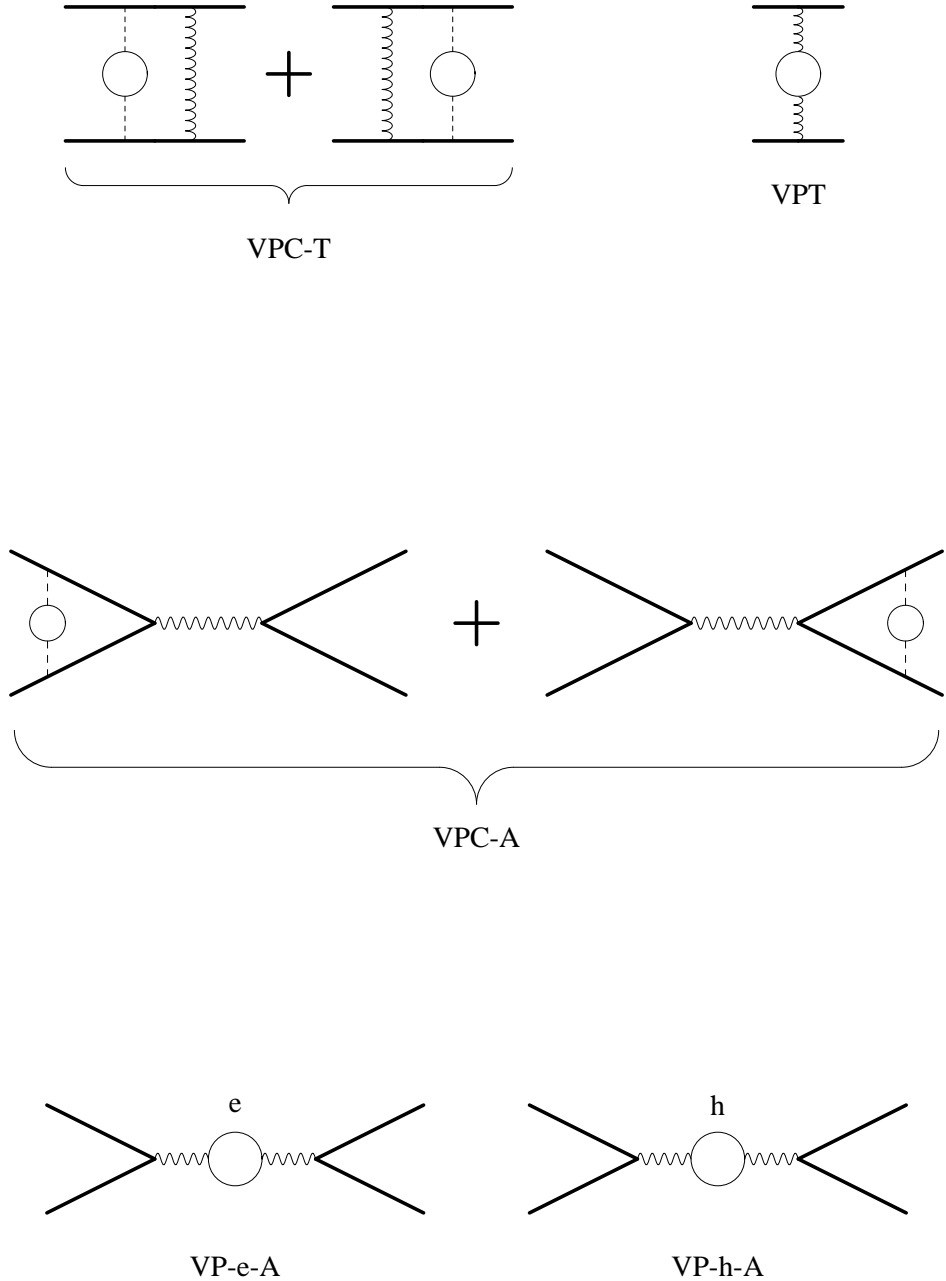}}}
\caption{\label{ours} Corrections of order $\alpha/\pi\,E_F$ to
the hyperfine splitting specific to the dimuonic atom. Diagrams are
further explained in the text.}
\end{figure}

Now we turn to the evaluation of corrections specific to dimuonium
(the relevant diagrams are depicted in Fig. \ref{ours}).
We first evaluate the correction due to vacuum polarization insertions
in the Coulomb photon line. 
For the dimuonic atom, the correction to the wave function due to
electronic vacuum polarization has relative order $\alpha/\pi$ and
therefore modifies the hyperfine splitting in $\alpha/\pi$ 
relative order.
The order of magnitude of this correction  
is specific to the dimuonic atom (small length scale of the system,
see Section \ref{intro}.  
For ``VPC-T'' (see Fig. \ref{ours}), the contribution 
is given by the matrix element
\begin{equation}
\Delta E_{\rm VPC-T}  = 2 \times \frac{4}{3} \, 
\frac{\pi \, \alpha}{m^2_{\mu}} \, 
\langle\psi|\delta({\bbox r})\,{\overline G}(E_{\psi})\,V_U|\psi\rangle,
\end{equation}
where 
\begin{equation}
\overline{G}(E_\psi) = \sum_{\psi_n \neq \psi} \, 
\frac{|\psi_n\rangle \langle\psi_n|}{E_\psi-E_n}
\end{equation}
is the reduced Coulomb Green function (the pole of
the reference state is excluded from the sum over all
intermediate states). Because atomic momenta are of the
order of $\alpha \, m_\mu$ in the system under consideration, it is
sufficient to carry out the calculations with non--relativistic 
wave functions. The reduced Green
function is a function of two coordinates ${\bbox r}_1$ and 
${\bbox r}_2$. Due to the appearance of the $\delta$-like potential 
the Green function is needed only for ${\bbox r}_1 = {\bbox 0}$.
It can be expressed in closed form \cite{IKe}, 
\begin{equation}
\overline{G}_{1S}(E_{1S}; 0, {\bf r}) =
\frac{\alpha m_r^2}{4\pi}\frac{2e^{-z_1/2}}{z_1}
\Big[2z_1(\ln{z_1}+C)+z_1^2-5z_1-2\Big],
\end{equation}
and
\begin{equation}
\overline{G}_{2S}(E_{2S}; 0, {\bf r}) =
-\frac{\alpha m_r^2}{4\pi}\frac{e^{-z_2/2}}{2z_2}
\Big[4z_2(z_2-2)(\ln{z_2}+C)+z_2^3-13z_2^2+6z_2+4\Big],
\end{equation}
where $C=0.5772\dots$ is the Euler 
constant, $z_n=(2\,\alpha\,m_r\,r)/n$ and 
$m_r$ denotes the reduced mass of
the system ($m_r=m_{\mu}/2$ for the dimuonic atom). 
The radial integration can be performed by
standard formulae. We finally obtain a one-dimensional 
integral over the $v$ parameter of the Uehling potential. For the
$1S$ ground state, we obtain
\begin{eqnarray}
\Delta E_{\rm VPC-T}(1S) &=& \frac{\alpha}{\pi}\,E_F\,
\int_0^1 dv\,\frac{4\, x^2\, (1-v^2/3)}{(2 + x\,\sqrt{1-v^2} )^3} \,
\times \nonumber\\
& & \times \left[2 + \frac{2}{7\, x} \, \frac{1}{\sqrt{1-v^2}} + 
\frac{3}{7} \, \sqrt{1-v^2} + \left(\frac{4}{7} + \frac{2}{7} \, x \,
\sqrt{1-v^2} \right) \, \ln\left(1+\frac{2}{x\,\sqrt{1-v^2}}\right)
\right] \nonumber\\
&=& \frac{\alpha}{\pi}\,(0.605)\,E_F,
\end{eqnarray}
where 
\begin{equation}
x = \alpha \frac{m_\mu}{m_e} = 1.50886\,.
\end{equation}
For $2S$, a analogous evaluation yields
\begin{equation}
\Delta E_{\rm VPC-T}(2S) =
\frac{\alpha}{\pi}\,(0.523)\cdot\frac{E_F}{8}\,.
\end{equation}
The analytical calculations were partially performed with 
the computer algebra system {\sc Mathematica} \cite{mathematica}.
The contribution due to the VPC-A diagram
can be evaluated by considering the relative contribution of these
diagrams to first order hfs ($4/7$ and $3/7$, respectively). 
An independent evaluation is done by utilization of a spectral
decomposition of the reduced Coulomb Green function.
If that representation of the Green function is chosen, 
then the calculation is closely related to \cite{IKmu}. In order to simplify
the calculation, we observe that the perturbation can be expressed as
a modification of the  
wave function at $r = 0$. The discrete
and continuous spectra give distinct contributions. 
The results of the respective terms (discrete and continuous 
spectrum) for the $1S$ and $2S$ states are
\begin{equation} 
\label{res1S}
\Delta E_{{\rm VPC-T}}(1S) = 
\frac{\alpha}{\pi}\,E_F\cdot  
2 \cdot \frac{4}{7}\,\left[
\frac{\Delta \psi_{1S}(0)}{\psi_{1S}(0)} \right]_{\mu^+\mu^-} 
= \frac{\alpha}{\pi} \, \left[ 0.04 + 0.56 \right] \cdot E_F\,,
\end{equation}
and
\begin{equation} 
\label{res2S}
\Delta E_{{\rm VPC-T}}(2S) =  
\frac{\alpha}{\pi}\,\frac{E_F}{8}\cdot 2 \cdot \frac{4}{7}\,
\left[ \frac{\Delta \psi_{2S}(0)}{\psi_{2S}(0)} \right]_{\mu^+\mu^-}
= \frac{\alpha}{\pi} \, \left[ -0.14 + 0.66 \right] \cdot 
  \frac{E_F}{8}\,.
\end{equation}
The continuous spectrum contribution (second numerical term in
Eq. (\ref{res1S}) and (\ref{res2S}) is very large. 
It can be understood in the following way. If one would
try to omit the energy of the intermediate state in the denominator of the
expression
\begin{eqnarray}
\Delta\psi ({\bbox 0}) &=&
\int d^3 r\, {\overline G}(E_{\psi},{\bbox 0} , {\bbox r}) V_U(r) \,
\psi({\bbox r}) \nonumber\\
&=& 
\int d^3 r\,\sum_{nl\neq\psi} \frac{\psi_{nl}({\bbox 0})\,\psi^*_{nl}
({\bbox r})}{E_\psi - E_n} \, V_U({\bbox r}) \, \psi({\bbox r}) \nonumber 
\end{eqnarray}
in the sense of $E_\psi - E_n \to E_\psi$, then the result should 
include a divergence in the sum over states. In view of the
equation $\sum_{nl\neq\psi} \psi_{nl}({\bbox 0})\,\psi^*_{nl}({\bbox r}) =
\delta({\bbox r}) - \psi({\bbox 0})\,\psi^*({\bbox r})$ 
(completeness of the spectum), this is seen as follows:
\begin{eqnarray}
& & \int d^3 r\,\sum_{nl} \frac{1}{E_\psi} \,
\psi_{nl}({\bbox 0})\,\psi^*_{nl}({\bbox r}) \, V_U({\bbox r}) \, 
  \psi({\bbox r}) \nonumber\\
&=& 
 \frac{ V_U({\bbox 0}) - 
  \langle \psi | \, V_U \, | \psi \rangle}{E_\psi} \,
\psi({\bbox 0}), \nonumber
\end{eqnarray}
and $V_U({\bbox 0})$ is a diverging quantitiy.
Its apperance means that virtual intermediate states with large 
wave numbers $k > \alpha \, m_\mu$ 
are important for the convergence of the integral for
$\Delta \psi ({\bbox 0})$, 
and they lead to a numerically 
large contribution.  The final results for the VPC-T and VPC-A
contributions are listed in Table \ref{alpha5}.

The correction due to the vacuum polarization insertion into 
the transverse photon line (VPT) is proportional to the matrix element
\begin{eqnarray}
\label{i-vpt}
{\cal M} &=& 
\langle \psi | {\bbox \nabla}^2 V_U | \psi \rangle \nonumber\\
&=& \frac{\alpha}{\pi} \, 
\int_0^1 dv\,\frac{v^2\,(1-v^2/3)}{1-v^2} \int d^3 r \,
\Big|\psi({\bbox r})\Big|^2\,
{\bbox \nabla}^2 \left[ \left(\frac{-\alpha}{r}\right)
\exp\left(-\frac{2\,m_e\, r}{\sqrt{1-v^2}}\right) \,
\right] \nonumber\\
&=& \frac{\alpha}{\pi} \, \int_0^1 dv\,\frac{v^2\,(1-v^2/3)}{1-v^2} \,
\int \frac{d^3p}{(2\pi)^3} \, \int \frac{d^3k}{(2\pi)^3} \,
\psi^*({\bbox p}) \,
\frac{- 4 \, \pi \,\alpha \,
({\bbox p} - {\bbox k})^2}{({\bbox p} - {\bbox k})^2 +
4\,m_e^2/(1-v^2)} \, \psi({\bbox k}).
\end{eqnarray} 
The matrix element ${\cal M}$ is evaluated both in momentum space 
and in coordinate space with the same result. We obtain for the $1S$ state
\begin{equation}
{\cal M}(1S) = \frac{\alpha^2\,\big(\alpha \, m_{\mu}\big)^4 }{2\pi} \, 
\int_0^1 dv\;\frac{v^2\,(1-v^2/3)}{1-v^2} \; 
\frac{ 2\lambda
+ \alpha \, m_{\mu}}{(\lambda  + \alpha\,m_{\mu})^2}\,,
\end{equation}
where $\lambda$ is defined in Eq. (\ref{deflambda}).
The final results for VPT (after numerical $v$-integration) are 
\begin{equation}
\Delta E_{\rm VPT}(1S) = \frac{\alpha}{\pi} \, 0.345 \cdot E_F \, \quad
\mbox{and} \, \quad 
\Delta E_{\rm VPT}(2S) = \frac{\alpha}{\pi} \, 0.355 \cdot E_F \,.
\end{equation}
The results are included in Table \ref{alpha5}.

The term due to the muonic vacuum polarization in the 
photon line in the annihilation diagram (denoted
``VP-$\mu$-A'') is known, since it has the
same relative magnitude as that of electronic vacuum polarization in
the positronium system. The electronic part (denoted as VP-e-A) 
is found from the well known asymptotic behavior of the vacuum 
polarization \cite{IV,IZ}
\begin{equation}  \label{1loopVP}
\Delta E_{{\rm VPeA}} (nS) = \frac{\alpha}{\pi}\,\frac{3}{7} \,
\left[\frac{1}{3}\ln{\frac{q^2}{m^2_e}}-\frac{5}{9} - 
\frac{\pi}{3} \, i\right] \cdot \frac{E_F}{n^3}
\end{equation}
for $q^2=(2\,m_\mu)^2$. The
contribution due to the electronic vacuum polarization in the annihilation is
the second largest correction to the dimuonium hyperfine splitting
in $\alpha/\pi$ relative units (see Table \ref{alpha5}, 
result for VP-e-A).
The time-like 
vector $q = (2\,m_{\mu}, {\bbox 0})$ is lying in the far 
time-like asymptotic region for the electronic vacuum polarization.
The logarithmic term which appeared originally by the replacement
$\ln{\big(\bbox{q}^2/m^2\big)} \to \ln{\big(-q^2/m^2\big)}$ 
(going from space--like to
time--like virtual photons) leads to the imaginary part.
It can be ascribed to the decay of the
ortho state into free electrons ($\mu^+\mu^- \to \gamma \to e^+e^-$)
\begin{equation}  
\label{Tortho}
\frac{1}{\Gamma^{(0)}(n^3S_1)} =
\tau^{(0)}(n^3S_1) =  
\frac{6\,n^3}{\alpha^5\,m_\mu}
=n^3\cdot 1.806 \cdot 10^{-12}\,{\rm s}\,.
\end{equation}
In contrast to positronium, the parastate and orthostate lifetime 
have the same order of magnitude in dimuonium ($\alpha^5\,m_\mu$). 

The hadronic contribution (denoted by VP-h-A) to the 
vacuum polarization
is found as the sum of four terms. The main
contribution results from a pionic loop. We follow here the approach 
in \cite{jon1}. The spectral function is of the form 
\begin{equation}
\rho(s) = \frac{(s - 4\,m_{\pi}^2)^{3/2}}{12\,s^{5/2}}\,
|F_{\pi}(s)|^2 \,,
\end{equation}
where the pionic form factor is used in the form given by Gounaris and
Sakurai \cite{fpi}
\begin{equation}  \label{SGff}
F_{\pi}(s) = \frac{N}{D_1 + D_2 - i\,D_3}\,.
\end{equation}
The quantities $N$, $D_1$, $D_2$ and $D_3$ are given by
\begin{equation}
N = m_\rho^2 + d m_\rho \, \Gamma_\rho\,,
\end{equation}
\begin{equation}
d = \frac{3}{\pi} \frac{m_\pi^2}{k_\rho^2} 
\ln\frac{m_\rho + 2 \, k_\rho}{2\,m_\pi} + 
\frac{m_\rho}{2\,\pi\,k_\rho} - 
\frac{m_\pi^2\, m_\rho}{\pi\,k_\rho^3} \approx 0.48,
\end{equation}
\begin{equation}
D_1 = m_\rho^2 - s, \quad D_3 = m_\rho\,\Gamma_\rho \, 
\Big(k(s)/k_\rho\Big)^3 \,
m_\rho/\sqrt{s},
\end{equation}
\begin{equation}
D_2 = \Gamma_\rho \frac{m_\rho^2}{k_\rho^3} \,
\bigg[k(s)^2\,\left(h(s) - h_\rho\right) + k_\rho^2 \,\, h'(m_\rho^2) \,\,
(m_\rho^2 - s) \bigg],
\end{equation}
where $h'$ denotes the derivative of $h$, 
and the functions $k$ and $h$ are defined as
\begin{equation}
k(s)=\frac{1}{2}\sqrt{s - 4m_\pi^2}, \quad h(s) = \frac{2}{\pi} \,
\frac{k(s)}{\sqrt{s}} \, 
\ln \left(\frac{\sqrt{s} + 2\, k(s)}{2\,m_\pi}\right)\,.
\end{equation}
with the special values $k_\rho \equiv k(m_\rho^2)$, $h_\rho \equiv
h(m_\rho^2), \Gamma_\rho = 150.7(1.2) \, {\rm MeV},
m_\rho = 768.5(6) \, {\rm MeV}$ \cite{pdg96}. 
We give results for the $1S$ state only in the sequel.
The $1/n^3$ scaling is easily restored in the final result. 
The contribution from the pionic vacuum polarization is given by
\begin{equation}
\Delta E_{\pi^{+}\pi^{-}}(1S) = 
\frac{\alpha}{\pi}\,\frac{3}{7}\,\left[4\,m_\mu^2\,
\int_{4\, m_\pi^2}^{\infty} ds\,\frac{\rho(s)}{4\, m_\mu^2-s}\right] 
\, E_F =  
\frac{\alpha}{\pi} \, (-0.055) \cdot E_F\,.
\end{equation}
In the simple $\rho$-meson pole approximation, where $\rho(s) =
4\pi^2/f_\rho^2 \, \delta(s - m_\rho^2)$ with $f_\rho^2/(4\,\pi) =
2.2$ \cite{bauer}, we have
\begin{equation}
\Delta E_{\pi^{+}\pi^{-}}(1S) \approx
\Delta E_{\rho}(1S) = \frac{\alpha}{\pi} \, (-0.050)\, E_F\,.
\end{equation}
This result agrees with the full pion form factor of 
(\ref{SGff}) to about $10\,\%$,
so it is justified to consider mesonic resonances of higher energy in
the pole approximation only.
The higher energy mesonic contributions are due to the $\omega$ and $\phi$
resonances. These resonances are not included in the Gounaris-Sakurai 
form factor and are treated separately. Estimating the coupling constants as
$f_\omega^2/(4\pi) = 18(2)$, $f_\phi^2/(4\pi) = 11(2)$ 
\cite{jon1,bauer} and given the meson
masses of $m_\omega = 782 \, {\rm MeV}$,  $m_\phi = 1019 \, {\rm
MeV}$ \cite{pdg96}, the results are
\begin{equation}
\Delta E_{\omega}(1S) = \frac{\alpha}{\pi} \, (-0.006) \cdot E_F
\quad\mbox{and}\quad
\Delta E_{\phi}(1S) = \frac{\alpha}{\pi} \, (-0.005) \cdot E_F\,.
\end{equation}
The background 
above $1 \, {\rm GeV}$ is estimated by assuming a form of
\begin{equation}
\rho(s) = \frac{R}{3\,s}, \quad \mbox{where} \quad
R= \frac{\sigma(e^{+}\,e^{-} \to {\rm hadrons})}
 {\sigma(e^{+}\,e^{-} \to \mu^{+}\,\mu^{-})}\,,
\end{equation}
for the spectral function with a branching ratio $R\approx 2$
(constant) below $\sqrt{s} = 4\,{\rm GeV}$ and $R\approx 4$ above 
$\sqrt{s} = 4\,{\rm GeV}$ (see \cite{pdg96}, p. 190). 
Integrating from an estimated lower threshold of 
$s_{\rm th} \approx (1 \, {\rm GeV})^2$ we obtain 
\begin{equation}
\Delta E_{>}(1S) \approx \frac{\alpha}{\pi} \, (-0.014) \cdot E_F.
\end{equation}
Summing all contributions and restoring the $1/n^3$ scaling,
we have as the contribution from the hadronic
vacuum polarization
\begin{equation}
\Delta E_{\rm VP-h-A} (nS) = 
\Delta E_{\pi} + \Delta E_{\omega} + \Delta
E_{\phi} + \Delta E_{>} = 
\frac{\alpha}{\pi} \, \left[-0.080(9)\right]\,\cdot\,\frac{E_F}{n^3}\,.
\end{equation}
We estimate model-dependent uncertainties in the hadronic vacuum
polarization to be of the order of $11 \, \%$.
The hadronic term is included in Table
\ref{alpha5} (result for VP-h-A).

The sum of all corrections (see Table \ref{alpha5})
to the hyperfine splitting in dimuonium amounts to
\begin{equation}
\Delta E_{\rm hfs}(1S) = \frac{\alpha}{\pi} \, 0.689(9) 
\cdot E_F
\quad \mbox{and} \quad
\Delta E_{\rm hfs}(2S) = \frac{\alpha}{\pi} \, 0.556(9) 
\cdot \frac{E_F}{8}\,.
\end{equation}
In the final results for hyperfine splitting, we estimate 
higher order corrections to enter at the $5\%$ level of the
next--to--leading order contributions. We obtain
\begin{equation}
E_{\rm hfs}(1S) = 4.23283(35) \cdot 10^7 \, {\rm MHz}
\end{equation}
and
\begin{equation}
E_{\rm hfs}(2S) = 5.28941(34) \cdot 10^6 \, {\rm MHz}\,.
\end{equation}

Results for the Lamb shift and the hyperfine structure of low-lying levels
are presented in Tables \ref{alpha3} and  
\ref{alpha5}. 
The largest deviation from the scaling appeared in VPC-T and VPC-A
contributions ($16\%$), because the wave function (see Eqs.
(\ref{res1S}) and (\ref{res2S})) is more sensitive to the behaviour
of the potential about the origin. 

An important point for a possible investigation of the 
spectrum is the lifetime of levels. As one
can see from Eqs. (\ref{Tpara}) and (\ref{Tortho}) the annihilation 
lifetimes of $S$-levels are much shorter that the lifetime of the
free muon, $\tau_\mu=2.20\, 10^{-6}\, s$. By contrast,
the annihilation decay rate for 
$P$-states includes an extra $\alpha^2$, and the lifetime 
is of the same order of magnitude as that of the free muon. 

The decay rates of excited states have to be compared with 
the atomic transition rates, which are also of 
the order of $\alpha^5\,m_\mu$ (see e. g. \cite{BS}).
We obtain
\begin{equation}
\tau(2P\to 1S) = 1.54 \cdot 10^{-11}\,{\rm s}\,.
\end{equation}
Annihilation of $P$ states is suppressed by two orders of
$\alpha$ compared to $S$ states due to the behaviour of the $P$
wave function near the origin. 
Therefore, the annihilation lifetime of the $2P$ state
can be estimated to be of the order of $10^{-7}\,{\rm s}$, and it is seen
that the atomic transition, not the annihilation, 
determines the lifetime of the excited $2P$
state in dimuonium. 

This situation is different for $S$ states, where the annihilation
dominates over atomic transitions.
The annihilation lifetimes of of $1S$ and $2S$ states lie between $0.6 \cdot
10^{-12}\, {\rm s}$ and $14 \cdot 10^{-12}\,{\rm s}$,
and we expect that the decay rates could be measured 
via detection of the decay products. 
Radiative corrections to the decay rates are considered in the 
following Section.

\section{DECAY CHANNELS OF ORTHODIMUONIUM}
\label{secOM}

The leading--order 
contributions to the orthodimuonium and paradimuonium decay 
rate can be extracted as the imaginary parts of 
the energy corrections to the hyperfine structure,
\begin{equation}
\label{leadingOrder}
\Gamma^{(0)}(n^3S_1) =
\frac{\alpha^5 \, m_\mu}{6\,n^3} \quad \mbox{and} \quad
\Gamma^{(0)}(n^1S_0) =
\frac{\alpha^5 \, m_\mu}{2\,n^3}\,.
\end{equation}
The calculations at leading order were 
presented above (Eqs. (\ref{Tpara}, \ref{Tortho})).
The above results can also be found in \cite{malenfant1,malenfant2}.

\begin{figure}[htb]
\centerline{\mbox{\epsfysize=18.0cm\epsffile{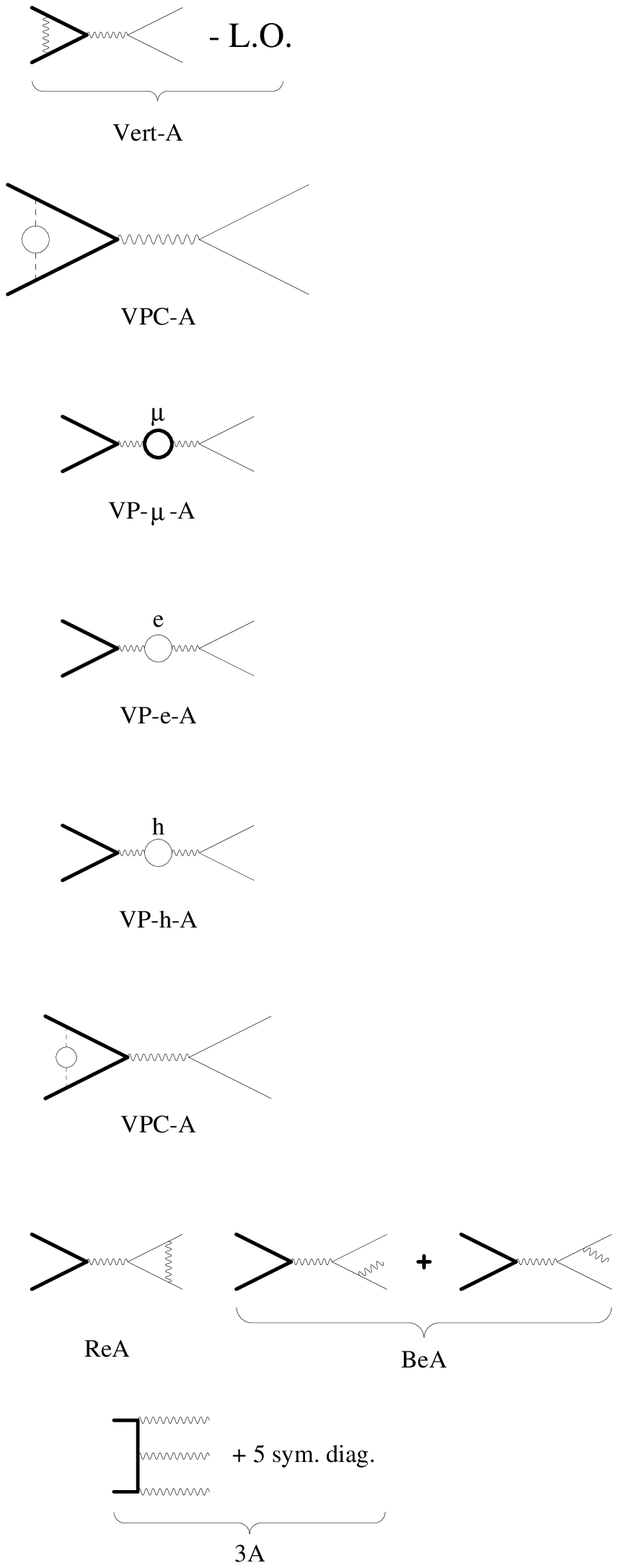}}}
\caption{\label{decayOM} Order $\alpha/\pi$
corrections to the decay channels of orthodimuonium. The 5 symmetrical
diagrams originate from the symmetrization of 
photon wave functions.}
\end{figure}

We begin the consideration of radiative corrections with 
orthodimuonium, and provide results for the $1S$ state here. 
Results for the $2S$ state (and $1S$) are summarized in Table \ref{alpha5O}.
The diagrams contributing in next--to--leading order for ortho states are 
depicted in Fig. \ref{decayOM}. 

The VPC-A correction
can be interpreted as a modification of the wave function at the
origin. The energy shift and decay rate are 
both proportional to $|\psi(\bbox{0})|^2$. The annihilation
diagram contributes $3/7\,E_F$ to the first--order result for
the hyperfine splitting. Hence, the VPC-A diagram
yields a correction of 
\begin{equation}
\Delta \Gamma_{\rm VPC-A}(1^3S_1) = \frac{\alpha}{\pi} \left[\frac{7}{3} \cdot 
0.454 \right]\, \Gamma^{(0)}(1^3S_1) = 
\frac{\alpha}{\pi} 1.06 \cdot \Gamma^{(0)}(1^3S_1)
\end{equation}
to the decay rate. Analogous considerations are true for the
Vert-A correction, because the diagrams consist of separated blocks,
and hence the correction to hfs and to the decay rate can be traced
back to the same matrix element of the wave function. We obtain
\begin{equation}
\Delta \Gamma_{\rm Vert-A}(n^3S_1) = - 4 \,
\frac{\alpha}{\pi} \cdot \Gamma^{(0)}(n^3S_1)\,.
\end{equation}

The diagrams VP-$\mu$-A, VP-e-A and VP-h-A have to be interpreted as
modifications of the photon propagator. Because the energy shift is
proportional to the amplitude of the propagator, 
but the decay rate is proportional to
its square, we have to multiply the relative correction to hfs by a
factor of 2 in order to obtain the relative correction to the decay 
rate. Hence, the VP-$\mu$-A, VP-e-A and VP-h-A diagrams
yield a total correction of
\begin{displaymath} 
\frac{\alpha}{\pi} \, \left[-\frac{16}{9} + \frac{4}{3} \, 
\ln\left(2 \frac{m_\mu}{m_e}\right) - 
\frac{10}{9} - 0.37(4) \right] \cdot 
\Gamma^{(0)}(n^3S_1)
\end{displaymath}
to the decay rate.

\begin{figure}[htb]
\centerline{\mbox{\epsfysize=10.0cm\epsffile{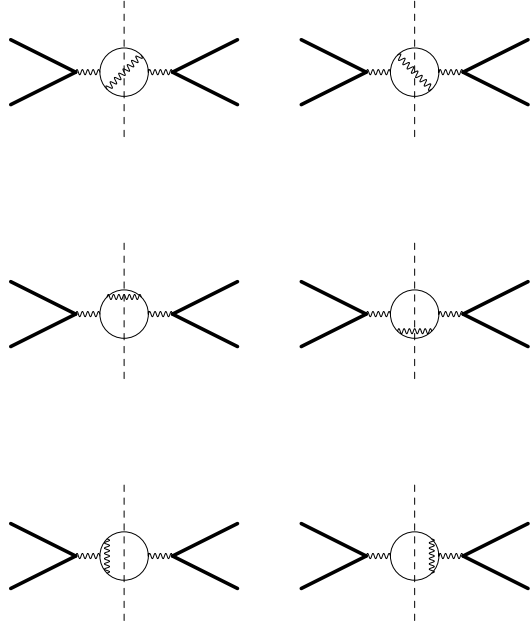}}}
\caption{\label{decayVPcutOM} Evaluation of bremsstrahlung 
and electron vertex corrections
to orthodimuonium decay as imaginary part of the two--loop 
vacuum polarization insertion in the photon line.}
\end{figure}

We now consider the one-loop 
radiative corrections to the electron 
line (ReA) and emission of a photon
by an electron, i. e. bremsstrahlung (BeA). The sum 
of BeA$+$ReA can be easily obtained from
the diagrams in Fig. \ref{decayVPcutOM} for the two-loop electron 
polarization correction to the hyperfine splitting,
which is completely determined by 
the asymptotic behaviour of the two-loop vacuum polarization
(see e. g. \cite{Schwinger}). For the correction to the hyperfine
splitting, we obtain
\begin{equation}
\Delta E_{{\rm VP-2}} (nS) = \frac{\alpha^2}{\pi^2} \,\frac{3}{7}\,
\left[\frac{1}{4}\ln{\frac{ q^2}{m^2_e}}+
\left(\zeta(3)-\frac{5}{24} \right)
-\frac{\pi}{4} \, i\right] \frac{1}{n^3} \cdot E_F\,,
\end{equation}
where $q^2=4m_\mu^2$ and $\zeta(3)=1.202\dots$ is the Riemann 
$\zeta$ function
of argument 3. This correction is of relative order $\alpha^2/\pi^2$
with respect to $E_F$ 
and was therefore not considered in Section \ref{hfs}.
The imaginary part of this contribution is just the result for the 
sum of the diagrams ReA$+$BeA. The correction to the decay rate can
now be evaluated easily. The sum of ReA$+$BeA yields a correction of
\begin{displaymath}
\frac{3}{4} \, \frac{\alpha}{\pi} \cdot \Gamma^{(0)}(n^3S_1)
\end{displaymath}
relative to the leading order result 
(cf. Eqs. (\ref{1loopVP},\ref{leadingOrder})). 

The last term in relative order 
$\alpha/\pi$ is the three-photon annihilation.
The result is known from orthopositronium 
calculations \cite{3phot} (see also \cite{BS,IZ}): 
\begin{equation}
\Delta \Gamma_{\rm 3A}(n^3S_1) =  \frac{2\alpha^6}{\pi} \,\frac{1}{n^3} \,
\frac{\pi^2-9}{9} \, m_\mu = \frac{\alpha}{\pi} \, 
\left[\frac{4}{3} \left(\pi^2 - 9\right)\right] \cdot 
\Gamma^{(0)}(n^3S_1)\,.
\end{equation}
The final result for the $1S$ decay (see also Table \ref{alpha5O}) is
\begin{eqnarray}
\label{OMalp}
\Delta \Gamma(1^3S_1) &=&  \frac{\alpha}{\pi} \,
\left[
\left(
\frac{4}{3}\ln{\frac{2\,m_\mu}{m_e}}-\frac{221}{36}
\right) 
+ \Big(0.68(4)\Big) +\left(\frac{4\,(\pi^2-9)}{3}
\right)
\right]  \, \Gamma^{(0)}(1^3S_1)\nonumber\\
&\approx& 
\frac{\alpha}{\pi} \,
\bigg\{ \Big(1.90\Big)
 + \Big(0.68(4)\Big) + \Big(1.16\Big)
\bigg\}  \, \Gamma^{(0)}(1^3S_1) \nonumber\\
&=& \frac{\alpha}{\pi} \,
3.74(4) \cdot \Gamma^{(0)}(1^3S_1)\,.
\end{eqnarray}
The first term corresponds to the sum of the analytically evaluated 
contributions Vert-A, VP-$\mu$-A, VP-e-A and BeA$+$ReA, and has
the numerical value $1.90$. 
The second term originates
from the numerically evaluated contributions VPC-A 
and VP-h-A. 
The last term is associated to the three photon decay. 
The result in Eq. (\ref{OMalp}) has to be compared with the earlier
analysis of the decay rate of heavy leptonium \cite{malenfant2}. 
The final result for the analytically evaluated
contributions, which is identical to
the first term in Eq. (\ref{OMalp}), whose analytical expression is 
$(4/3)\,\ln(2\,m_\mu/m_e) - 221/36$, is found in agreement
with the pioneering investigations by J. Malenfant \cite{malenfant2}. 
For the
VPC-A correction, Malenfants results are in slight
numerical disagreement
with ours (see Eqs. (77,78) in \cite{malenfant2}). We presume this 
disagreement can be traced to the fact that Malenfant has 
calculated the
VPC-A correction with free Green functions, whereas the evaluation
in this work is done using bound Green functions. In this context
it is important to note that in the limit of 
$\alpha\,m_\mu/2m_e \to 0$, our result is agreement with that
of Malenfant. This can be seen as follows. The VPC-A correction
may be rewritten as a correction $\Delta \psi(0)$ to the wave function at the
origin. After this reinterpretation, we find
\begin{equation}
\Delta \psi(0) = 
  \frac{\alpha}{\pi}\,\left(\frac{3\,\pi}{16}\,\kappa + 
     O(\kappa^2) \right)\,\psi(0)
\quad\mbox{for}\quad \kappa = \frac{\alpha\,m_\mu}{2\,m_e} \to 0\,,
\end{equation}
which is in agreement with Eq. (80) of \cite{malenfant2}.
We can therefore conclude that in the limit of weak binding
($\kappa\to0$), Malenfants result is in agreement with ours.
However, we hold the view that
bound Green functions should rather be used for the VPC-A
correction. The atomic
momentum in dimuonium is of the order
of $\alpha \, m_\mu/2 = \kappa\, m_e$ ($\kappa \approx 0.75$).
This momentum is close to
the mass of the loop particles (electrons and positrons) 
of electronic vacuum polarization.
These particles determine the radius of the Uehling potential.
Therefore, some of the momentum integration 
for the VPC-A
correction is performed in an area about $m_e$, 
where the bound Coulomb Green function cannot be approximated
by the free Green function (because the effect of
the binding Coulomb potential, in momentum space, is
inversely proportional to the square of the momentum transfer).
This consideration should explain the slight numerical disagreement
for the VPC-A correction between this work and the result in
\cite{malenfant2}.

The sum of next--to--leading order corrections to the decay rate 
for the $2S$ state is (see Table \ref{alpha5O})
\begin{equation}
\Delta \Gamma(2^3S_1) = 
\frac{\alpha}{\pi} \,
3.60(4) \cdot \Gamma^{(0)}(2^3S_1)\,.
\end{equation}
We estimate the higher order corrections to be suppressed by an
additional factor of $\alpha$ compared to the next--to--leading order
result. We obtain the following results for the decay rate of orthodimuonium,
taking into account also the uncertainty from our model of the hadronic
vacuum polarization:
\begin{equation}
\tau(1^3S_1) = 1.79073(77) \cdot 10^{-12}\,{\rm s} \quad\mbox{and}\quad
\tau(2^3S_1) = 14.3305(59) \cdot 10^{-12}\,{\rm s}.
\end{equation}

\section{DECAY CHANNELS OF PARADIMUONIUM}
\label{secPM}

\begin{figure}[htb]
\centerline{\mbox{\epsfysize=14.0cm\epsffile{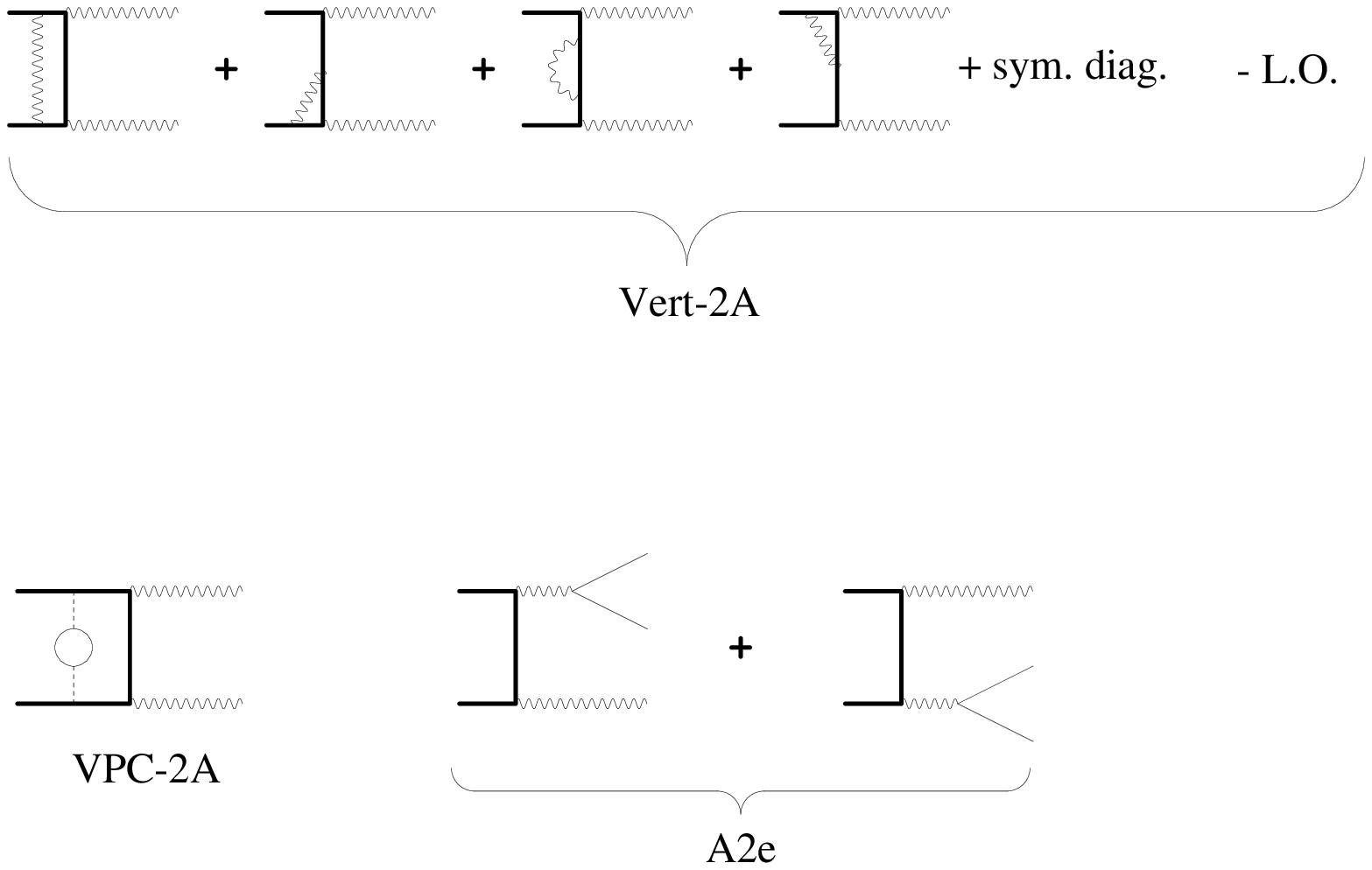}}}
\caption{\label{decayPM} Order $\alpha/\pi$
corrections to the decay channels of paradimuonium. The symmetrical 
diagrams originate from the symmetrization of the photon wave 
functions.}
\end{figure}

The diagrams contributing to paradimuonium decay 
in next--to--leading order $\alpha^6\, m$ are presented
in Fig \ref{decayPM}, results are summarized in Table \ref{alpha5P}.
The vertex correction  term 
(Vert-2A) is equivalent to the corresponding correction
for parapositronium \cite{alphaP},
\begin{equation}
\Delta \Gamma_{\rm Vert-2A}(n^1S_0) =  -\frac{\alpha}{\pi} \,
\frac{20-\pi^2}{4} \, \Gamma^{(0)}(n^1S_0)\,.
\end{equation}
The correction to the wave function caused by the VPC-A diagram
modifies the decay rate of the para system in the same way as
the ortho system (see Table \ref{alpha5P}), and introduces a deviation
from the $1/n^3$ scaling.

\begin{figure}[htb]
\centerline{\mbox{\epsfysize=10.0cm\epsffile{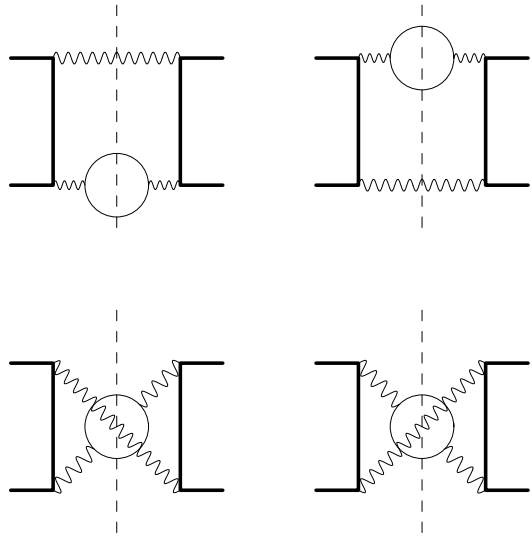}}}
\caption{\label{decayVPcutPM} Evaluation of corrections to 
the paradimuonium decay caused by the production of an electron positron 
pair (A2e correction).}
\end{figure}

For the para state, there exists another correction to the decay rate 
corresponding to the decay into a photon and an electron-positron pair.
The result for the A2e correction can be obtained in the following way. 
We consider the one--loop vacuum polarization insertion
into a 2-photon annihilation diagram (Fig. \ref{decayVPcutPM}). 
Because the vacuum
polarization insertion, evaluated for a real photon, 
must be equal to zero (gauge invariance 
of vacuum polarization), 
the imaginary part of the diagrams in Fig. \ref{decayVPcutPM} leads to the result
we need. We can use Eq. (\ref{vpe}) for the parametric form
of the vacuum polarization
insertion, and we consider the $s$-integration 
as the final one. The integrand is now
equivalent to the imaginary part 
of the 2A diagram for the hfs, 
but with one of the photons having a finite
mass $s$ ($s$ corresponds to the sum of the
four--momenta of the emerging electron--positron pair). 
The vacuum polarization insertion normally 
fixes a gauge (the Landau gauge) for
the virtual photon, because the polarization 
insertion is proportional
to the transverse projector. However, as it was 
demonstrated in \cite{mugauge},
it is possible to substitute in Eq. (\ref{vpe})
any covariant gauge. For convenience, we choose the Feynman gauge.
As a result we have the expression
\begin{equation}
\Delta \Gamma_{\rm A2e}(n^1S_0) =  
\frac{\alpha}{\pi} \,\int ds \, \rho(s) \, 
\Gamma^{(0)}(s,0) \,,
\end{equation}
where $\Gamma^{(0)}(s,0)$ is the decay rate to 
one real photon and 
a virtual photon with mass $s$. In order to obtain the
correction in relative units, we divide by the first--order result,
which is given by $\Gamma^{(0)}(0,0)$. The correction relative to the
first--order result $\Gamma^{(0)}(n^1S_0)$ is given by
\begin{equation}
\frac{\Delta \Gamma_{\rm A2e}(n^1S_0)}{\Gamma^{(0)}(n^1S_0)} 
= 2\,\frac{\alpha}{\pi} \,\int ds\,\rho(s)\,
\frac{\Gamma^{(0)}(s,0)}{\Gamma^{(0)}(0,0)}\,,
\end{equation}
An additional factor 2 appears because the
insertion of a vacuum polarization operator doubles the number of
non-equivalent diagrams contributing to the imaginary part.
For the logarithmic coefficient we may neglect $s$ in
$\Gamma^{(0)}{(s,0)}$ and approximate
$\Gamma^{(0)}{(s,0)} \to \Gamma^{(0)}{(0,0)}$, 
and use the asymptotic form of the spectral
function (cf. Eq (\ref{defrho})),
\begin{displaymath}
\rho(s) \to \frac{1}{3\,s} \quad \mbox{for} \quad s \to \infty\,.
\end{displaymath}
We can thus easily obtain the logarithmic coefficient,
\begin{displaymath}
\frac{\Delta \Gamma_{\rm A2e}(n^1S_0)}{\Gamma^{(0)}(n^1S_0)} 
\approx 2\,\int_{(2\,m_e)^2}^{\Lambda^2 = (2\,m_\mu)^2}
ds \, \frac{1}{3\,s} 
= 2\, \frac{1}{3}\,\ln\frac{m_\mu^2}{m_e^2} =
\frac{4}{3}\,\ln\frac{m_\mu}{m_e} \,.
\end{displaymath}
The full result requires a more detailed analysis of the dependence
of $\Gamma^{(0)}{(s,0)}$ on $s$. It differs from the approximate
analysis presented above only by an additive constant.
The final result of the calculation is  
\begin{equation}
\label{A2e}
\Delta \Gamma_{\rm A2e}(n^1S_0) =  \frac{\alpha}{\pi} \,
\left(\frac{4}{3}\ln{\frac{2m_\mu}{m_e}}-\frac{16}{9}
\right) \, \frac{1}{n^3} \cdot
 \Gamma^{(0)}(n^1S_0)\,.
\end{equation}
An independent evaluation of the A2e-correction using the standard
S-matrix formalism is used to verify the result in Eq. (\ref{A2e}).
Treatment of the Dirac currents involved in simplified by
application of a symbolic program \cite{hip} developed
for high energy physics calculations by A. Hsieh and E. Yehudai. 
Some care must be taken during evaluation, because one cannot 
assume the electrons as massless in the final states (the result else
includes a logarithmic divergence in the electron mass). Proper 
regularization of the relevant expression then leads to the result
in Eq. (\ref{A2e}).

The final result for the next--to--leading order radiative corrections 
to paradimuonium decay (see Table \ref{alpha5P}) is
\begin{equation}
\Delta \Gamma(1^1S_0) =  \frac{\alpha}{\pi} \,
4.79 \cdot \Gamma^{(0)}(1^1S_0)\,. 
\end{equation}
For the $2S$ state, we have
\begin{equation}
\Delta \Gamma(2^1S_0) =  \frac{\alpha}{\pi} \,
4.65 \cdot \Gamma^{(0)}(2^1S_0)\,. 
\end{equation}
Estimating higher order corrections to enter at the level of $5\%$ of the
next--to--leading order contributions, 
we obtain the following theoretical values
for the paradimuonium decay:
\begin{equation}
\tau(1^1S_0) = 0.59547(33) \cdot 10^{-12}\,{\rm s} \quad\mbox{and}\quad
\tau(2^1S_0) = 4.7653(25) \cdot 10^{-12}\,{\rm s}\,.
\end{equation}

\section{Conclusions}

We evaluate next--to--leading order corrections 
to the spectrum, to the hyperfine splitting and to the
decay rate of low-lying levels of the dimuonic system. The results 
for the spectrum are given in Section \ref{spectrum}. We observe that 
for $2P$ states, the atomic decay into the $1S$ state dominates over 
annihilation processes. This would facilitate experimental
observation of the atomic transition, if dimuonium atoms can be produced
in quantities sufficient to carry out spectroscopic measurements. 
We evaluate the hyperfine
splitting of $1S$ and $2S$ states in next--to--leading order. The
results are
\begin{equation}
E_{\rm hfs}(1S) = \left[1 + \frac{\alpha}{\pi} \, 0.689(9)\right] \,
\frac{7}{12}\,\alpha^4\,m_\mu 
=  4.23283(35) \cdot 10^7 \, {\rm MHz}
\end{equation}
and
\begin{equation}
E_{\rm hfs}(2S) = \left[1 + \frac{\alpha}{\pi} \, 0.556(9)\right] \,
\frac{7}{12}\,\alpha^4\,m_\mu 
= 5.28941(34) \cdot 10^6 \, {\rm MHz}\,.
\end{equation}

We present a complete evaluation of all next--to--leading order
radiative corrections to the lifetime of both the ortho-- and
para--state of the dimuonic atom. In leading order, we reproduce the 
known results \cite{malenfant1,malenfant2}
\begin{equation}
\Gamma(n^3S_1) = \frac{\alpha^5 \, m_\mu}{6\,n^3} \quad \mbox{and}
\quad
\Gamma(n^1S_0) = \frac{\alpha^5 \, m_\mu}{2\,n^3}
\end{equation}
as imaginary contributions to the hyperfine splitting
(Eqs. (\ref{Tpara}, \ref{Tortho})).

The results in next--to--leading order are for orthodimuonium,
\begin{equation}
\Gamma(1^3S_1) = 
\left[1 + \frac{\alpha}{\pi} \,
3.74(4) \right] \Gamma^{(0)}(1^3S_1)
\quad\mbox{and}\quad 
\Gamma(2^3S_1) = 
\left[1 + \frac{\alpha}{\pi} \,
3.60(4) \right] \Gamma^{(0)}(2^3S_1)\, ,
\end{equation}
where the primary theoretical uncertainty is due to hadronic vacuum
polarization. 
The lifetimes of orthostates are given by
\begin{equation}
\tau(1^3S_1) = 1.79073(77) \cdot 10^{-12}\,{\rm s} \quad\mbox{and}\quad
\tau(2^3S_1) = 14.3305(59) \cdot 10^{-12}\,{\rm s}\,.
\end{equation}
For paradimuonium, we obtain
\begin{equation}
\Gamma(1^1S_0) = 
\left[1 + \frac{\alpha}{\pi} \,
4.79 \right] \Gamma^{(0)}(1^1S_0)
\quad\mbox{and}\quad 
\Gamma(2^1S_0) = 
\left[1 + \frac{\alpha}{\pi} \,
4.65 \right] \Gamma^{(0)}(2^1S_0)\,.
\end{equation}
The lifetimes are given by
\begin{equation}
\tau(1^1S_0) = 0.59547(33) \cdot 10^{-12}\,{\rm s} \quad\mbox{and}\quad
\tau(2^1S_0) = 4.7653(25) \cdot 10^{-12}\,{\rm s}\,.
\end{equation}
We estimate higher order QED corrections to enter at the $5\%$ level of
the next--to leading order corrections considered in this work.

Lifetimes in the $10^{-12}\,{\rm s}$ range 
can be measured by established methods of particle physics
via detection of the decay products
(electron-positron pairs in case of 
orthodimuonium and two photons in the case
of paradimuonium). We stress that accurate decay rate measurements
can be accomplished with fewer individual atoms than would be needed
for spectroscopic investigations. 

One of the ways to investigate the hyperfine structure 
of $1S$ or $2S$ states
could be based on the observation 
of the interference between paradimuonium and 
orthodimuonium in an infrared frequency field at resonance 
(the radiofrequency field would mix the two states
and thus yield a modified decay rate of the statistical sample).
We conclude that the dimuonic system offers the possibility
to observe quantum electrodynamic effects in a previously unexplored
kinematic region. 

\section*{Acknowledgments}

G. S. and U. J. thank 
DFG for continued support (contract no. SO333/1-2).
The work of S. K. and V.~I. has been 
supported in part by the Russian
Foundation for Basic Research (grant $\#$95-02-03977).
S. K. is grateful for 
hospitality at the Technical 
University of Dresden. 
The authors would like to thank V. V. Vereshagin, V. A. Shelyuto,
P. Mohr and J. Malenfant for 
stimulating discussions. We also acknowledge helpful discussions
with M. Sander and S. Vigdor.

\newpage

%
%

\begin{table}[htb]
\begin{center}
\begin{minipage}{7cm}
\begin{tabular}{ccc}
&&\\[-1ex]
state & $C$ & ${\cal L}$ 
\\[1ex]
\hline
\hline
&&\\[-1ex]
 1S & -0.15 & -0.49 {\rm eV}   \\[1ex]
 2S & -0.018 & -0.058 {\rm eV}   \\[1ex]
 2P & -.000043 & -0.0014 {\rm eV}   \\[1ex]
\end{tabular}
\end{minipage}
\end{center}
\caption{\label{alpha3} Contributions to the 
Lamb shift due to the electronic vacuum
polarization. $C$ is given in relative units, 
$\Delta E = \frac{\alpha}{\pi} \, C \, E_0$, where  
$E_0$ is the Rydberg constant for the dimuonic 
atom.} 
\end{table}

\newpage

%
%

\begin{table}[htb]
\begin{center}
\begin{minipage}{10cm}
\begin{tabular}{ccc}
&&\\[-1ex]
diagram & $C(1S)$ & $8\,C(2S)$ 
\\[1ex]
\hline
\hline
&&\\[-1ex]
(g-2)-T & 0.571 & 0.571   \\[1ex]
Rec & -0.857  &  -0.857   \\[1ex]
Vert-A & -1.714 & -1.714   \\[1ex]
VP-$\mu$-A & -0.381 & -0.381   \\[1ex]
2A & 0.263 &  0.263   \\[1ex]
\hline
&&\\[-1ex]
VPC-T & 0.605 & 0.523   \\[1ex] 
VPT & 0.345 & 0.355   \\[1ex]
VPC-A & 0.454 & 0.393   \\[1ex]
VP-e-A & 1.483 & 1.483   \\[1ex]
VP-h-A & -0.080(9) & -0.080(9)
\\[1ex]
\hline
\hline
&&\\[-1ex]
Sum & 0.689(9) & 0.556(9) \\[1ex]
\end{tabular}
\end{minipage}
\end{center}
\caption{\label{alpha5} Corrections of relative order
$\alpha/\pi$ to the dimuonium hyperfine splitting. All contributions are
explained in the text. The corrections (g-2)-T, Rec, Vert-A,
VP-$\mu$-A and 2A contribute to the positronium hyperfine splitting just 
as in dimuonium. The remaining contributions (VPC-T, VPT, VPC-A,
VP-e-A, VP-h-A) are specific to the
dimuonic system. Results are given in relative units   
$\Delta E = \alpha/\pi \, C \, E_F$. }
\end{table}

\newpage

%
%

\begin{table}[htb]
\begin{center}
\begin{minipage}{10cm}
\begin{tabular}{ccc}
&&\\[-1ex]
diagram & $C(1S)$ & $8\,C(2S)$ 
\\[1ex]
\hline
\hline
&&\\[-1ex]
Vert-A & -4.00 & -4.00 \\[1ex]
VPC-A & 1.06 & 0.92 \\[1ex]
VP-$\mu$-A & -1.78 & -1.78 \\[1ex]
VP-e-A & 6.92 & 6.92 \\[1ex]
VP-h-A & -0.37(4) & -0.37(4) \\[1ex]
BeA$+$ReA & 0.75 & 0.75 \\[1ex]
3A & 1.16 & 1.16 \\[1ex]
\hline
\hline
&&\\[-1ex]
Sum & 3.74(4) & 3.60(4) \\[1ex]
\end{tabular}
\end{minipage}
\end{center}
\caption{\label{alpha5O} Corrections of relative order
$\alpha/\pi$ to the orthodimuonium decay. Results are 
given in relative units   
$\Delta \Gamma(n^3S_1) = \alpha/\pi \, C \, \Gamma^{(0)}(1^3S_1)$. }
\end{table}

\newpage

%
%

\begin{table}[htb]
\begin{center}
\begin{minipage}{10cm}
\begin{tabular}{ccc}
&&\\[-1ex]
diagram & $C(1S)$ & $8\,C(2S)$ 
\\[1ex]
\hline
\hline
&&\\[-1ex]
Vert-2A & -2.53 & -2.53 \\[1ex]
VPC-2A & 1.06 & 0.92   \\[1ex]
A2e & 6.26 & 6.26  \\[1ex]
\hline
\hline
&&\\[1ex]
Sum & 4.79  & 4.65 \\[1ex]
\end{tabular}
\end{minipage}
\end{center}
\caption{\label{alpha5P} Corrections of relative order
$\alpha/\pi$ to the paradimuonium decay. Results are given in relative units   
$\Delta \Gamma(n^1S_0) = \alpha/\pi \, C \, \Gamma^{(0)}(1^1S_0)$. }
\end{table}

\newpage

%
%

%
%

%
%

%
%

%
%

%
%

\begin{figure}[htb]
\centerline{\mbox{\epsfxsize=12.0cm\epsffile{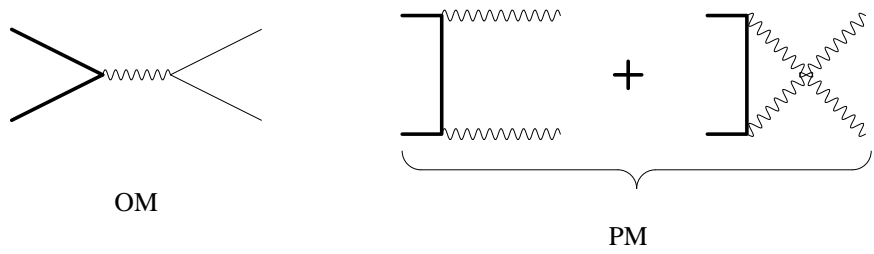}}}
\caption{\label{decay} Decay channels of orthodimuonium (one--photon
decay into an electron positron pair, OM) and paradimuonium 
(two photon decay, PM). The decay channels depicted yield the main
contribution to the decay of the system ($\Gamma$ of
order $\alpha^5 \, m_\mu$). Bold fermionic lines denote 
muons, thin lines denote electrons and positrons. }
\end{figure}

%
%

%
%

%
%

%
%

\end{document}